\documentclass[aps,prd,twocolumn,groupedaddress,amsmath,amssymb]{revtex4-1}
\usepackage{graphicx}  
\usepackage{dcolumn}   
\usepackage{bm}        
\usepackage{verbatim}   
\usepackage[utf8]{inputenc}
\usepackage{color}

\begin{document}

\title{Polarization of prompt $J/\psi$ and $\Upsilon$(1S) production in the color evaporation model}
\author{Vincent Cheung}
\affiliation{
   Department of Physics,
   University of California, Davis,
   Davis, CA 95616, USA
   }
\author{Ramona Vogt}
\affiliation{
   Nuclear and Chemical Sciences Division,
   Lawrence Livermore National Laboratory,
   Livermore, CA 94551, USA
   }
\affiliation{
   Department of Physics,
   University of California, Davis,
   Davis, CA 95616, USA
   }
\date{\today}
\begin{abstract}
We calculate the polarization of prompt $J/\psi$ and $\Upsilon$(1S) production using the color evaporation model at leading order. We present the polarization parameter $\lambda_\vartheta$ as a function of center of mass energy and rapidity in $p+p$ collisions. We also compare the $x_F$ dependence to experimental results in $p$+Cu and $\pi$+W collisions, and predict the $x_F$ dependence in $p$+Pb collisions at fixed-target energies. At energies far above the $Q\overline{Q}$ production threshold, we find the prompt $J/\psi$ and $\Upsilon$(1S) production to be longitudinally polarized with $\lambda_\vartheta^{J/\psi}=-0.51^{+0.05}_{-0.16}$ and $\lambda_\vartheta^{\Upsilon \rm{(1S)}}=-0.69^{+0.03}_{-0.02}$. Both prompt $J/\psi$ and prompt $\Upsilon$(1S) are also longitudinally polarized at central rapidity, becoming transversely polarized at the most forward rapidities.
\end{abstract}

\pacs{
14.40.Pq
}
\keywords{
Heavy Quarkonia}

\maketitle


\section{Introduction}
One of the best ways to understand hadronization in QCD is to study the production of quarkonium. However, the production mechanism of quarkonium is still uncertain. Nonrelativistic QCD (NRQCD) \cite{Caswell:1985ui}, the most widely used model for quarkonium prduction encounters serious challenges in both the universality of the long distance matrix elements (LDMEs) and prediction of quarkonium polarization. The production cross sections in NRQCD, based on an expansion in the strong coupling constant and the $Q\overline{Q}$ velocity \cite{Bodwin:1994jh}, is factorized into hard and soft contributions and divided into different color and spin states. The LDMEs, which weight the contributions from each color and spin state, are fit to the data above some minumum transverse momentum, $p_T$. These LDMEs, which are conjectured to be universal, fail to describe the yields and polarization simultaneouly for $p_T$ cuts less than twice the mass of the quarkonium state \cite{Bodwin:2014gia,Faccioli:2014cqa}. They also depend on the collision system \cite{
Ma:2010yw,Butenschoen:2010rq,Gong:2012ug,Zhang:2009ym}. Moreoever, the polarization predicted by NRQCD is senstive to the $p_T$ cut. Thus the LDMEs are not universal as conjectured. The $\eta_c$ $p_T$ distributions calculated with LDMEs obtained from $J/\psi$ yields using heavy quark spin symmetry \cite{HQSS1,HQSS2,HQSS3}, overshoots the high $p_T$ LHCb $\eta_c$ results \cite{Butenschoen:2014dra} in a recent analysis. The color evaporation model (CEM) and NRQCD can describe production yields rather well but spin-related measurements like the polarization are strong tests of production models.

The CEM \cite{Barger:1979js,Barger:1980mg,Gavai:1994in,Ma:2016exq}, which considers all $Q\overline{Q}$ ($Q$ = $c$, $b$) production regardless of the quark color, spin, and momentum, is able to predict both the total yields and the rapidity distributions with only a single normalization parameter \cite{NVF}. We have previously presented the first polarization results in the CEM \cite{Cheung:2017loo}, which only considered charmonium and bottomonium production in general. This paper serves as a continuation of the previous work by presenting a leading order (LO) CEM calculation of the polarization in prompt $J/\psi$ and $\Upsilon$(1S) production.  It is still a $p_T$-independent result because there are no exclusive NLO polarized $Q\overline{Q}$ calculations on which to impose the $H\overline{H}$ ($H$ = $D$, $B$) mass threshold. Our calculation is another step toward a full CEM polarization result that provides a general idea of whether there is any appreciable LO polarization that might carry through to the next order even though the kinematics are different. We will begin to address the $p_T$ dependence in a subsequent publication.

In the traditional CEM, all quarkonium states are treated the same as $Q\overline{Q}$ below the $H\overline{H}$ threshold where the invariant mass of the heavy quark-antiquark pair is restricted to be less than twice the mass of the lowest mass meson that can be formed with the heavy quark as a constituent. The distributions for all quarkonium family members are assumed to be identical. In this paper, we use an improved CEM (ICEM) \cite{Ma:2016exq} where the invariant mass of the intermediate heavy quark-antiquark pair is constrained to be larger than the mass of produced quarkonium state, $M_Q$, instead of using the same lower limit of integration in the traditional CEM, $2m_Q$, as in our previous work and in Ref.~\cite{Barger:1979js}. The improved CEM describes the charmonium yields as well as the ratio of $\psi^\prime$ over $J/\psi$ better than the traditional CEM. In a $p+p$ collision, the production cross section for a quarkonium state is then
\begin{eqnarray}
\label{cem_sigma}
\sigma &=& F_Q \sum_{i,j} \int^{4m_H^2}_{M_Q^2}d\hat{s} \int dx_1 dx_2 f_{i/p}(x_1,\mu^2) f_{j/p}(x_2,\mu^2) \nonumber \\
&\times& \hat{\sigma}_{ij}(\hat{s}) \delta(\hat{s}-x_1 x_2 s) \;,
\end{eqnarray}
where $i$ and $j$ are $q$, $\overline{q}$ and $g$ such that $ij = q\overline{q}$ or $gg$. The square of the heavy quark pair invariant mass is $\hat{s}$ while the square of the center-of-mass energy in the $p+p$ collision is $s$. Here $f_{i/p}(x,\mu^2)$ is the parton distribution function (PDF) of the proton as a function of the fraction of momentum carried by the colliding parton $x$ at factorization scale $\mu$ and $\hat{\sigma}_{ij}$ is the parton-level cross section. Finally, $F_{Q}$ is a universal factor for the quarkonium state and is independent of the projectile, target, and energy. At leading order, the rapidity distribution, $d\sigma/dy$, in the ICEM is
\begin{eqnarray}
\label{cem_rapidity}
\frac{d\sigma}{dy} &=& F_{Q} \sum_{i,j} \int^{4m_H^2}_{M_Q^2} \frac{d\hat{s}}{s}  f_{i/p}(x_1,\mu^2)f_{\overline{q}/p}(x_2,\mu^2) \nonumber \\
&\times& \hat{\sigma}_{ij} (\hat{s})  \;,
\end{eqnarray}
where $x_{1,2} = (\sqrt{\hat{s}/s}) \exp(\pm y)$. The longitudinal momentum fraction distribution, $d\sigma/dx_F$, in the ICEM is 
\begin{eqnarray}
\label{cem_xf}
\frac{d\sigma}{dx_F} &=& F_{Q} \sum_{i,j} \int^{2m_H}_{M_Q} \frac{d\sqrt{\hat{s}}}{s} \frac{2 \sqrt{\hat{s}}}{\sqrt{x_F^2 + 4\hat{s}/{s}}} \nonumber \\
&\times&f_{i/p}(x_1,\mu^2)f_{j/p}(x_2,\mu^2) \hat{\sigma}_{ij} (\hat{s}) \;,
\end{eqnarray}
where $x_{1,2} = (\pm x_F+\sqrt{x_F^2+4\hat{s}/s})/2$. We take the square of the factorization and renormalization scales to be $\mu^2 = \hat{s}$.


\section{Polarization of directly produced $Q\overline{Q}$}
\begin{figure}
\centering
\includegraphics[width=\columnwidth]{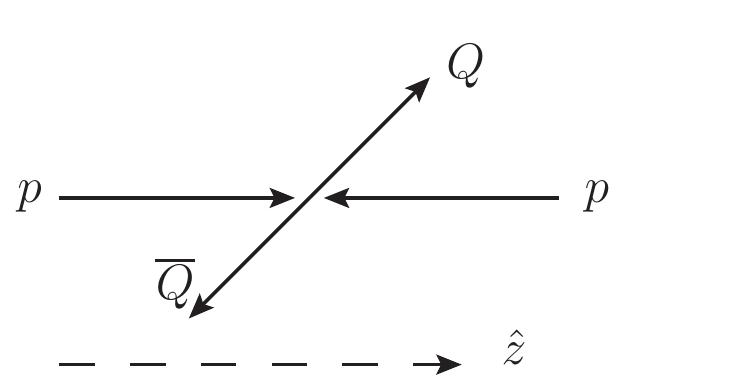}
\caption{\label{polarization} The orientation of the $z$-axis is indicated by the dashed arrowed line. Two proton arrows indicate the incoming beam directions. If the quarks in the final state heavy quark-antiquark pair have the same helicity, then the total angular momentum along the $z$-axis, $J_z$, is 0 while if they have opposite helicity, then $J_z = \pm 1$.}
\end{figure}

At leading order in $\alpha_s$, the final state $Q\overline{Q}$ pair is produced with zero total transverse momentum. We define the polarization axis ($z$-axis) in the helicity frame pointing from $\overline{Q}$ to $Q$ along the beam axis as shown in Fig.~\ref{polarization}.

There are four $\mathcal{O}(\alpha_s^2)$ Feynman diagrams to consider, one for $q\overline{q} \rightarrow Q\overline{Q}$ and three for $gg \rightarrow Q\overline{Q}$. Each diagram includes a color factor $C$ and a scattering amplitude $\mathcal{A}$. The generic matrix element for each process can be written as \cite{Cvitanovic:1976am}
\begin{eqnarray}
\mathcal{M}_{qq} &=& C_{qq} \mathcal{A}_{qq} \;, \\
\mathcal{M}_{gg} &=& C_{gg,\hat{s}} \mathcal{A}_{gg,\hat{s}} + C_{gg,\hat{t}} \mathcal{A}_{gg,\hat{t}} + C_{gg,\hat{u}} \mathcal{A}_{gg,\hat{u}} \;.
\end{eqnarray}
As previously mentioned, there is one diagram only for $q\overline{q} \rightarrow Q\overline{Q}$, thus a single amplitude, $\mathcal{A}_{qq}$. However, there are three diagrams for $gg \rightarrow Q \overline Q$ at leading order, the 
$\hat{s}$, $\hat{t}$ and $\hat{u}$ channels. In terms of the Dirac spinors $u$ and $v$, the individual amplitudes are
\begin{eqnarray}
\mathcal{A}_{qq} &=& \frac{g_s^2}{\hat{s}} [\overline{u}(p^\prime) \gamma_\mu v(p)][\overline{v}(k) \gamma^\mu u(k^\prime)] \;, \\
\mathcal{A}_{gg,\hat{s}} &=& - \frac{g_s^2}{\hat{s}} \Big\{ -2k^\prime \cdot \epsilon(k) [\overline{u}(p^\prime) \epsilon\!\!\!/(k^\prime) v(p)] \nonumber \\
&+&2 k\cdot \epsilon(k^\prime) [\overline{u}(p^\prime) \epsilon\!\!\!/(k) v(p)] \nonumber \\
&+& \epsilon(k) \cdot\epsilon(k^\prime)[\overline{u}(p^\prime) (k\!\!\!/^\prime - k\!\!\!/) v(p)]\Big\} \;, \\
\mathcal{A}_{gg,\hat{t}} &=& -\frac{g_s^2}{\hat{t}-M^2} \overline{u}(p^\prime) \epsilon\!\!\!/(k^\prime) (k\!\!\!/ -p\!\!\!/ +M) \epsilon\!\!\!/(k) v(p) \;, \\
\mathcal{A}_{gg,\hat{u}} &=& -\frac{g_s^2}{\hat{u}-M^2} \overline{u}(p^\prime) \epsilon\!\!\!/(k) (k\!\!\!/^\prime -p\!\!\!/ +M) \epsilon\!\!\!/(k^\prime) v(p) \;.
\end{eqnarray}
Here $g_s$ is the gauge coupling, $M$ is the mass of heavy quark ($m_c$ for charm and $m_b$ for bottom), $\epsilon$ represents the gluon polarization vectors, $\gamma^\mu$ are the gamma matrices, $k^\prime$ ($k$) is the momentum of initial state light quark (antiquark) or gluon, and $p^\prime$ ($p$) is the momentum of final sate heavy quark (antiquark). 

At leading order, the final state $Q\overline{Q}$ is produced with no dependence on the azimuthal angle and thus $L_z=0$. To extract the projection on a state with orbital-angular-momentum quantum number $L$, we find the corresponding Legendre component $\mathcal{A}_L$ in the amplitudes by
\begin{eqnarray}
\mathcal{A}_{L=0} &=& \frac{1}{2} \int_{-1}^{1} dx \mathcal{A}(x=\cos \theta) \;, \\
\mathcal{A}_{L=1} &=& \frac{3}{2} \int_{-1}^{1} dx \; x \mathcal{A}(x=\cos \theta) \;.
\end{eqnarray}

The final state total spin is determined by the helicities of the heavy quarks. Two helicity combinations that results in $S_z=0$ are added and normalized to give contribution to the spin triplet state ($S=1$). Having the amplitudes for $S=1$ with $S_z = 0,\pm 1$, and $L=0,1$ with $L_z = 0$, we calculate the amplitudes for $J=0,1,2$. First, the amplitudes for $J=1$, obtained by adding $S=1$ and $L=0$, are simply:
\begin{eqnarray}
\mathcal{A}_{J=1,J_z=\pm1} &=& \mathcal{A}_{L=0,L_z=0;S=1,S_z=\pm 1} \;, \\
\mathcal{A}_{J=1,J_z=0} &=& \mathcal{A}_{L=0,L_z=0;S=1,S_z=0} \;.
\end{eqnarray}
Then, using angular momentum algebra, the amplitudes for $J=0,1,2$, found by adding $S=1$ and $L=1$, are:
\begin{eqnarray}
\mathcal{A}_{J=0,J_z=0} &=& -\sqrt{\frac{1}{3}} \mathcal{A}_{L=1,L_z=0;S=1,S_z=0} \;, \\
\mathcal{A}_{J=1,J_z=\pm1} &=& \mp \frac{1}{\sqrt{2}} \mathcal{A}_{L=1,L_z=0;S=1,S_z=\pm 1} \;, \\
\mathcal{A}_{J=1,J_z=0} &=& 0 \;, \\
\mathcal{A}_{J=2,J_z=\pm2} &=& 0 \;, \\
\mathcal{A}_{J=2,J_z=\pm1} &=& \frac{1}{\sqrt{2}} \mathcal{A}_{L=1,L_z=0;S=1,S_z=\pm1} \;, \\
\mathcal{A}_{J=2,J_z=0} &=& \sqrt{\frac{2}{3}} \mathcal{A}_{L=1,L_z=0;S=1,S_z=0} \;.
\end{eqnarray}

\begin{table}
\caption{\label{states}The mass $M_Q$, the feed down contribution ratio $c_Q$, and the squared feed down transition Clebsch-Gordan coefficients $S_Q^{J_z}$ for all quarkonium states contributing to the prompt production of $J/\psi$ and $\Upsilon$(1S). We assume the $c_Q$ for $\chi_{b1}$(1P) and $\chi_{b2}$(1P) to be equal as well as that for $\chi_{b1}$(2P) and $\chi_{b2}$(2P).}
\begin{ruledtabular}
\begin{tabular}{ccccc}
$Q$ & $M_Q$ (GeV) & $c_{Q}$ & $S_Q^{J_z=0}$ & $S_Q^{J_z=\pm1}$\\
\hline
$J/\psi$ & 3.10 & 0.62 & 1 & 0 \\
$\psi$(2S) & 3.69 & 0.08 & 1 & 0 \\
$\chi_{c1}$(1P) & 3.51 & 0.16 & 0 & 1/2 \\
$\chi_{c2}$(1P) & 3.56 & 0.14 & 2/3 & 1/2 \\
$\Upsilon$(1S) & 9.46 & 0.52 & 1 & 0 \\
$\Upsilon$(2S) & 10.0 & 0.1 & 1 & 0 \\
$\Upsilon$(3S) & 10.4 & 0.02 & 1 & 0 \\
$\chi_{b1}$(1P) & 9.89 & 0.13 & 0 & 1/2 \\
$\chi_{b2}$(1P) & 9.91 & 0.13 & 2/3 & 1/2 \\
$\chi_{b1}$(2P) & 10.3 & 0.05 & 0 & 1/2 \\
$\chi_{b2}$(2P) & 10.3 & 0.05 & 2/3 & 1/2 \\
\end{tabular}
\end{ruledtabular}
\end{table}

Here, we have dropped terms that contain amplitudes of non-zero $L_z$. The amplitudes sorted by final state $J$ and $J_z$ are then squared while averaging over the polarization of the initial gluons or the spin of the light quarks, depending on the process, in the spirit of the CEM.

The squared matrix elements, $|\mathcal{M}|^2$, are calculated for each $J$, $J_z$ state. The color factors, $C$, are calculated from the SU(3) color algebra and are independent of final state angular momentum \cite{Cvitanovic:1976am}. They are
\begin{eqnarray}
|C_{qq}|^2 = 2 \;, |C_{gg,\hat{s}}|^2 = 12 \;, \nonumber \\ 
|C_{gg,\hat{t}}|^2 = \frac{16}{3} \;, |C_{gg,\hat{u}}|^2 = \frac{16}{3} \; .
\end{eqnarray}
\begin{eqnarray}
C_{gg,\hat{s}}^*C_{gg,\hat{t}} = +6 \;, C_{gg,\hat{s}}^*C_{gg,\hat{u}} = -6 \;, \nonumber \\
C_{gg,\hat{t}}^*C_{gg,\hat{u}} = -\frac{2}{3} \;.
\end{eqnarray}
Finally, the total squared amplitudes for a given $J,J_z$ state,
\begin{eqnarray}
|\mathcal{M}_{qq}^{J,J_z}|^2 &=&  |C_{qq}|^2 |\mathcal{A}_{qq}|^2 \;, \\
|\mathcal{M}_{gg}^{J,J_z}|^2 &=&  |C_{gg,\hat{s}}|^2 |\mathcal{A}_{gg,\hat{s}}|^2 + |C_{gg,\hat{t}}|^2 |\mathcal{A}_{gg,\hat{t}}|^2 \nonumber \\
&+& |C_{gg,\hat{u}}|^2 |\mathcal{A}_{gg,\hat{u}}|^2 + 2  C_{gg,\hat{s}}^*C_{gg,\hat{t}} \mathcal{A}_{gg,\hat{s}}^*\mathcal{A}_{gg,\hat{t}} \nonumber \\
&+& 2 C_{gg,\hat{s}}^*C_{gg,\hat{u}} \mathcal{A}_{gg,\hat{s}}^*\mathcal{A}_{gg,\hat{u}} \nonumber \\ 
&+& 2 C_{gg,\hat{t}}^*C_{gg,\hat{u}} \mathcal{A}_{gg,\hat{t}}^*\mathcal{A}_{gg,\hat{u}} \;,
\end{eqnarray}
are then used to obtain the partonic cross sections by integrating over solid angle:
\begin{eqnarray}
\hat{\sigma}_{ij}^{J,J_z} = \int d\Omega \Big( \frac{1}{8\pi} \Big)^2 \frac{|\mathcal{M}_{ij}^{J,J_z}|^2}{\hat{s}} \sqrt{1-\frac{4M^2}{\hat{s}}} \; .
\end{eqnarray}

The partonic cross sections for $J^P=1^-$ with $J_z = 0, \pm 1$ are found by adding the $L=0$ and $S=1$ contributions:
\begin{eqnarray}
\hat{\sigma}_{q\overline{q}}^{J_z = 0} (\hat{s}) &=& 0 \;, \\
\hat{\sigma}_{q\overline{q}}^{J_z = \pm 1}  (\hat{s}) &=& \frac{\pi \alpha_s^2}{9 \hat{s}} \chi \;, \\
\hat{\sigma}_{gg}^{J_z = 0} (\hat{s}) &=& \frac{7 \pi \alpha_s^2}{48\hat{s}} \frac{M^2}{\hat{s} \chi} \Big( \ln \frac{1+\chi}{1-\chi} \Big)^2 \;, \\
\hat{\sigma}_{gg}^{J_z = \pm 1} (\hat{s}) &=& \frac{\pi^3 \alpha_s^2}{1536\hat{s}} \chi \frac{(\sqrt{\hat{s}}-2M)(37\sqrt{\hat{s}}+38M)}{(2M+\sqrt{\hat{s}})^2} \;.
\end{eqnarray}
Here and in the following, $\chi = \sqrt{1-4M^2/\hat{s}}$.

The partonic cross sections for $J^P=0^+$, obtained by adding the $L=1$ and $S=1$ states, are
\begin{eqnarray}
\hat{\sigma}_{q\overline{q}}^{J_z = 0} (\hat{s}) &=& 0 \;, \\
\hat{\sigma}_{gg}^{J_z = 0} (\hat{s}) &=& \frac{9 \pi \alpha_s^2}{16\hat{s}} \frac{M^2}{\hat{s} \chi^3} \Big( 2\chi - \ln \frac{1+\chi}{1-\chi} \Big)^2 \;.
\end{eqnarray}

The individual partonic cross section for $J^P=1^+$ with $J_z = 0, \pm 1$, found by adding the contributions from $L=1$ and $S=1$, are
\begin{eqnarray}
\label{chi_1_qq}
\hat{\sigma}_{q\overline{q}}^{J_z = 0} (\hat{s}) &=& 0 \;, \\
\hat{\sigma}_{q\overline{q}}^{J_z = \pm 1}  (\hat{s}) &=& \frac{\pi \alpha_s^2}{18 \hat{s}} \chi \;, \\
\label{chi_1_gg}
\hat{\sigma}_{gg}^{J_z = 0} (\hat{s}) &=& 0 \;, \\
\hat{\sigma}_{gg}^{J_z = \pm 1} (\hat{s}) &=& \frac{3\pi^3 \alpha_s^2}{256\hat{s}} \chi \frac{(\sqrt{\hat{s}}-2M)(4\hat{s}-9M^2)}{(2M+\sqrt{\hat{s}})^3} \;.
\end{eqnarray}

The partonic cross sections for $J^P=2^+$ with $J_z = 0, \pm 1$, obtained by adding the $L=1$ and $S=1$ states, are
\begin{eqnarray}
\hat{\sigma}_{q\overline{q}}^{J_z = 0} (\hat{s}) &=& 0 \;, \\
\hat{\sigma}_{q\overline{q}}^{J_z = \pm 1}  (\hat{s}) &=& \frac{\pi \alpha_s^2}{18 \hat{s}} \chi \;, \\
\hat{\sigma}_{gg}^{J_z = 0} (\hat{s}) &=& \frac{9 \pi \alpha_s^2}{8\hat{s}} \frac{M^2}{\hat{s} \chi^3} \Big( 2\chi - \ln \frac{1+\chi}{1-\chi} \Big)^2 \;, \\
\hat{\sigma}_{gg}^{J_z = \pm 1} (\hat{s}) &=& \frac{3\pi^3 \alpha_s^2}{256\hat{s}} \chi \frac{(\sqrt{\hat{s}}-2M)(4\hat{s}-9M^2)}{(2M+\sqrt{s})^3} \;,
\end{eqnarray}
The sum of these results for each final state total angular momentum, $\sum_{J_z=-J}^{J_z=+J} \hat{\sigma}_{ij}^{J_z}$, is equal to the unpolarized partonic cross section $\hat{\sigma}_{ij}^{\rm unpol.}$.

Having computed the polarized $Q\overline{Q}$ production cross section at the parton level, we then convolute the partonic cross sections with the parton distribution functions (PDFs) to obtain the hadron-level cross section $\sigma$ as a function of $\sqrt{s}$ using Eq.~(\ref{cem_sigma}), and the rapidity distribution, $d\sigma/dy$, using Eq.~(\ref{cem_rapidity}). The quarkonium masses which appear as the lower limit of the $Q\overline{Q}$ invariant mass are listed in Table~\ref{states}. We employ the CTEQ6L1 \cite{Pumplin:2002vw} PDFs in this calculation and the running coupling constant $\alpha_s = g_s^2/(4\pi)$ is calculated at the one-loop level appropriate for the PDFs. 

\begin{figure}
\centering
\includegraphics[width=\columnwidth]{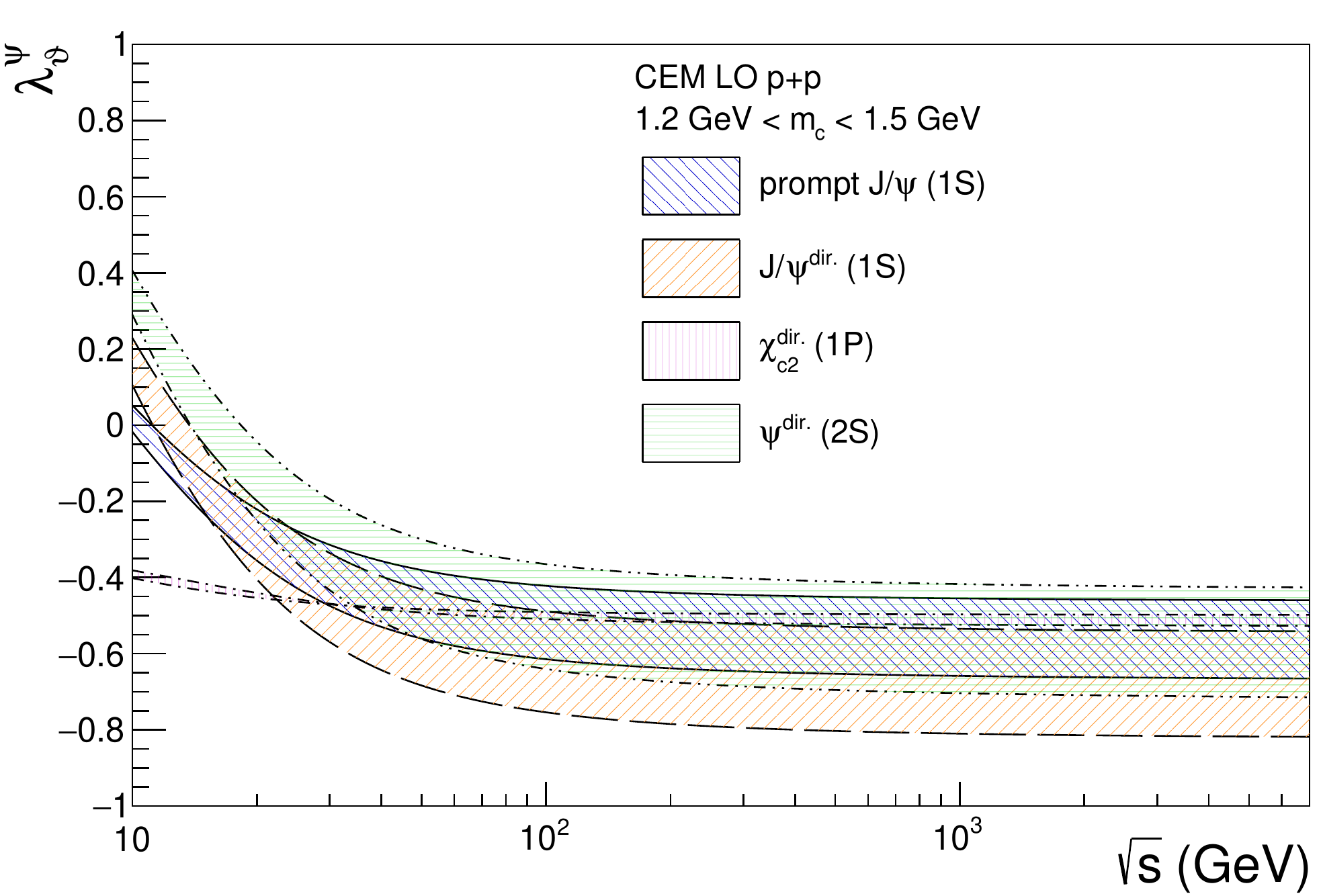}
\caption{\label{psi_lambda} The energy dependence of the polarization parameter $\lambda_\vartheta$ for production of prompt $J/\psi$ (solid), direct $J/\psi$ (dashed), direct $\chi_{c2}$(1P) (dot-dashed), and direct $\psi$(2S) (dot-dot-dashed).}
\end{figure}

\section{Polarization of prompt $J/\psi$ and $\Upsilon$(1S)}
We assume that the angular momentum of each directly produced quarkonium state is unchanged by the transition from the parton level to the hadron level, consistent with the CEM that the linear momentum is unchanged by hadronization.  This is similar to the assumption made in NRQCD that once a $c \overline c$ is produced in a given spin state, it retains that spin state when it becomes a $J/\psi$.

We calculate the $J_z=0,\pm1$ to unpolarized ratios for each directly produced quarkonium state $Q$ that has contribution to the prompt production of $J/\psi$ and $\Upsilon$(1S): $J/\psi$, $\psi$(2S), $\chi_{c1}$(1P), $\chi_{c2}$(1P), and $\Upsilon$(1S), $\Upsilon$(2S), $\Upsilon$(3S), $\chi_{b1}$(1P), $\chi_{b2}$(1P), $\chi_{b1}$(2P), $\chi_{b2}$(2P). These ratios, $R_{Q}^{J_z}$, are then independent of $F_Q$. We assume the feed down production of $J/\psi$ and $\Upsilon$(1S) from the higher mass bound state follows the angular momentum algebra. Their contributions to the $J_z=0$ to unpolarized ratios of prompt $J/\psi$ and $\Upsilon$(1S) are added and weighed by the feed down contribution ratios $c_{\psi}$ and $c_{\Upsilon}$ \cite{Digal:2001ue}:
\begin{eqnarray}
\label{mix_psi}
R_{J/\psi}^{J_z=0} &=& \sum_{\psi,J_z} c_{\psi} S_{\psi}^{J_z} R_{\psi}^{J_z} \;, \\
\label{mix_upsilon}
R_{\Upsilon\rm{(1S)}}^{J_z=0} &=& \sum_{\Upsilon,J_z} c_{\Upsilon} S_{\Upsilon}^{J_z} R_{\Upsilon}^{J_z} \;,
\end{eqnarray}
where $S_{Q}^{J_z}$ is the transition probability from a given state $Q$ produced in a given $J_z$ state to $J/\psi$ or $\Upsilon$(1S) with $J_z=0$ in a single decay. We assume two pions are emmited for an S state feed down, and a photon is emitted for a P state feed down. $S_{Q}^{J_z}$ is then 1 (if $J_z=0$) or 0 (if $J_z=1$) for $Q=\psi$(2S),$\Upsilon$(2S),$\Upsilon$(3S) since their transitions $Q \rightarrow J/\psi$+$\pi \pi$ or $Q \rightarrow \Upsilon$(1S)+$\pi \pi$ does not change the angular momentum. For directly produced $J/\psi$ and $\Upsilon$(1S), $S_{Q}^{J_z}$ is then 1 for $J_z=0$ and 0 for $J_z=1$. $S_{Q}^{J_z}$ for $\chi$ states are the squares of the Clebsch-Gordan coefficients for the feed down production from state $\chi$ to $J/\psi+\gamma$ or $\Upsilon$(1S)$+\gamma$.  The values of $M_Q$, $c_Q$ and $S_Q^{J_z}$ for all quarkonium states contributing to the prompt production of $J/\psi$ and $\Upsilon$(1S) are collected in Table~\ref{states}. We further assume that the contributions from $\chi_{b1}$(1P) and $\chi_{b2}$(1P) are the same, and also that the contributions from $\chi_{b1}$(2P) and $\chi_{b2}$(2P) are the same, similar to that in direct $J/\psi$ production.

\begin{figure}
\centering
\includegraphics[width=\columnwidth]{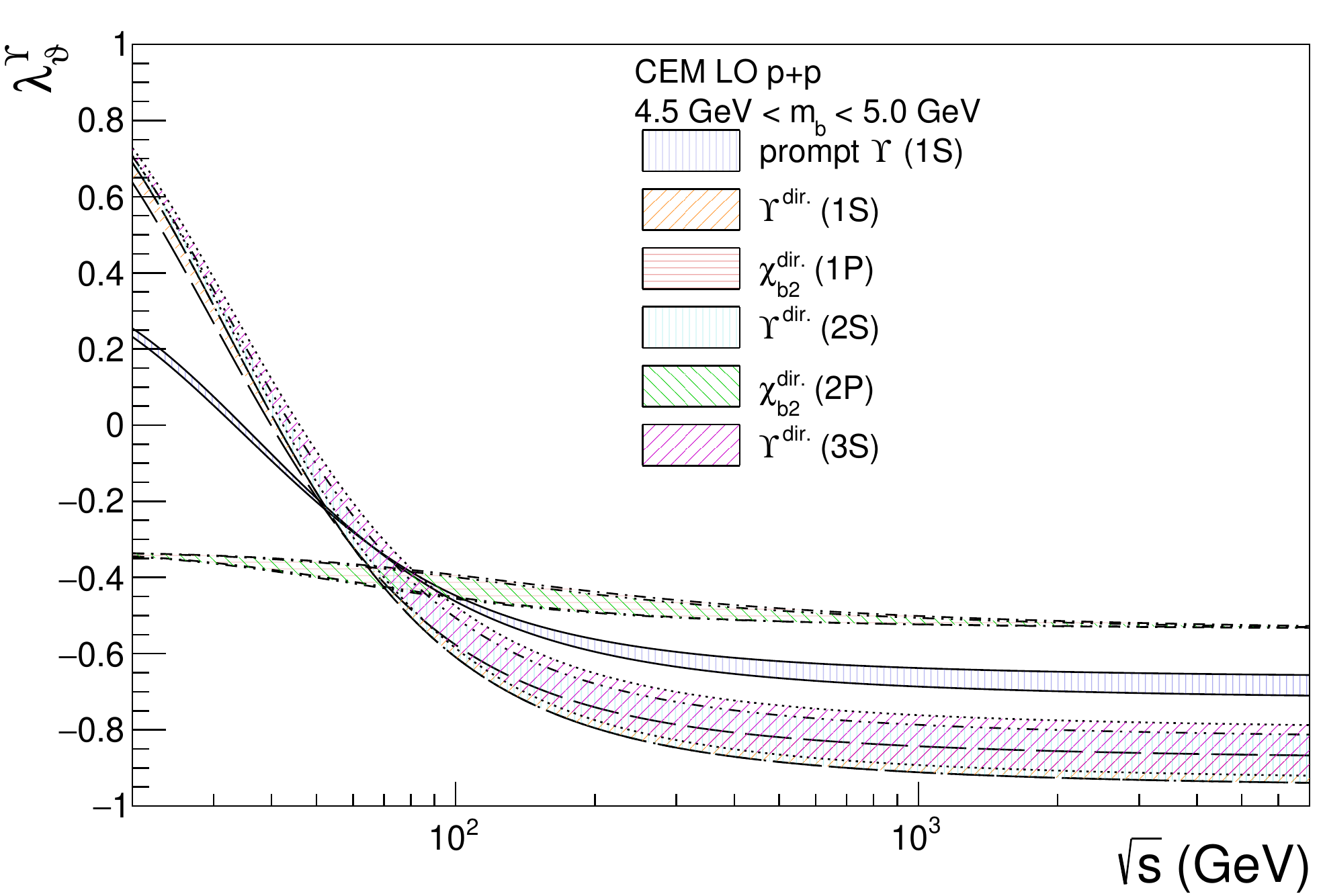}
\caption{\label{upsilon_lambda} The energy dependence of the polarization parameter $\lambda_\vartheta$ for production of prompt $\Upsilon$(1S) (solid), direct $\Upsilon$(1S) (dashed), direct $\chi_{b2}$(2P) (dot-dashed), direct $\Upsilon$(2S) (dot-dot-dashed), direct $\chi_{b2}$(2P) (dot-dot-dot-dashed), and direct $\Upsilon$(3S) (dotted). The result is shown for $\sqrt{s} > 20$~GeV to be above the $B\overline B$ threshold.}
\end{figure}

Finally, for each of the $J^{P}=1^-$ S states, the $J_z=0$ to unpolarized ratio is then converted into the polarization parameter $\lambda_\vartheta$ by \cite{Faccioli:2010kd}
\begin{eqnarray}
\label{s_state_lambda}
\lambda_{\vartheta} &=& \frac{1-3R^{J_z=0}}{1+R^{J_z=0}} \;.
\end{eqnarray}
Likewise, for the $J^{P}=1^+$ $\chi_1$P states, the $J_z=0$ to unpolarized ratio is converted into the polarization parameter $\lambda_\vartheta$ by \cite{Faccioli:2017nfn}
\begin{eqnarray}
\label{chi_1_lambda}
\lambda_{\vartheta} &=& \frac{-1+3R^{J_z=0}}{3-R^{J_z=0}} \;.
\end{eqnarray}
Also, for each of the $J^{P}=2^+$ $\chi_2$P states, the $J_z=0$ to unpolarized ratio is converted into the polarization parameter $\lambda_\vartheta$ by \cite{Faccioli:2017nfn}
\begin{eqnarray}
\label{chi_2_lambda}
\lambda_{\vartheta} &=& \frac{-3-3R^{J_z=0}}{9+R^{J_z=0}} \;.
\end{eqnarray}


\section{Results}

Since this calculation is LO in $\alpha_s$, we can only calculate the polarization parameter $\lambda_\vartheta$ as a function of $\sqrt{s}$ and $y$ (or $x_F$) but not $p_T$, which will require us to go to NLO, $\mathcal{O}(\alpha_s^3)$. However, the charm rapidity distribution at LO is similar to that at NLO \cite{Vogt:1995zf}. The same is true for $J/\psi$ production in the CEM. The only difference would be a rescaling of the parameter $F_Q$ based on the ratio NLO/LO using the NLO scale determined in Ref.~\cite{NVF}.  The unpolarized CEM results are in rather good agreement with the data from $p+p$ collisions \cite{NVF}.  

In the remainder of this section, we discuss the energy dependence of the polarization parameter $\lambda_\vartheta$ for the prompt production of $J/\psi$ and $\Upsilon$(1S), and direct production of quarkonium states that contribute to the feed down production. We then show the polarization parameter for prompt $J/\psi$ and $\Upsilon$(1S) production as a function of rapidity for selected energies. We also compare our results as a function of longitudinal momentum fraction to the polarization measured in fixed-target experiments as well as giving predictions for future fixed-target experiments. Finally, we discuss the sensitivity of our results to the choice of proton parton density functions, the factorization scale and the feed down ratios considered.

\begin{figure}
\centering
\includegraphics[width=\columnwidth]{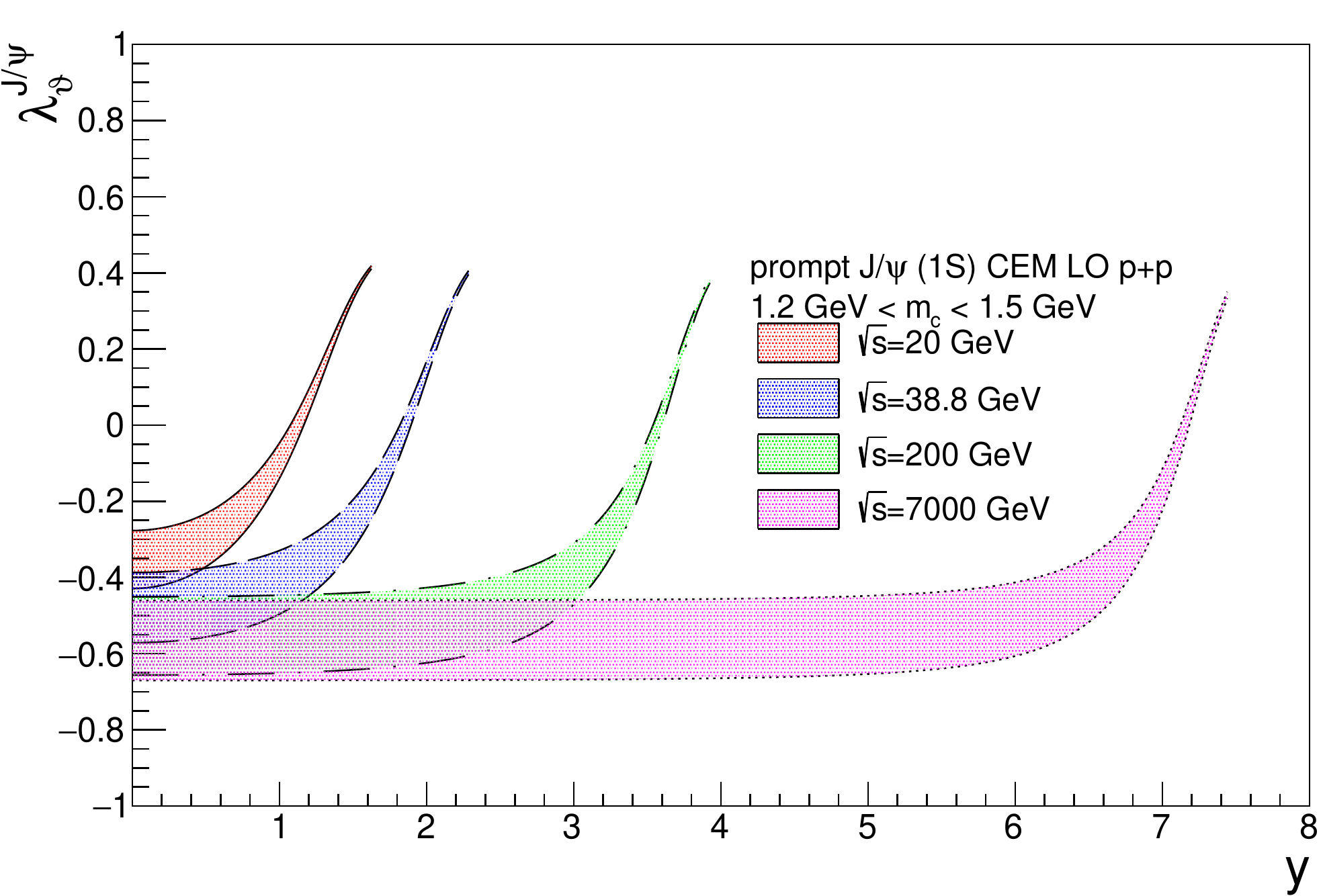}
\caption{\label{psi_lambda_y} The rapidity dependence of the polarization parameter $\lambda_\vartheta$ for production of prompt $J/\psi$ at $\sqrt{s}=20$~GeV (solid), 38.8~GeV (dashed), 200~GeV (dot-dashed), and 7000~GeV (dotted). The distributions are symmetric around $y=0$.}
\end{figure}

\begin{figure}
\centering
\includegraphics[width=\columnwidth]{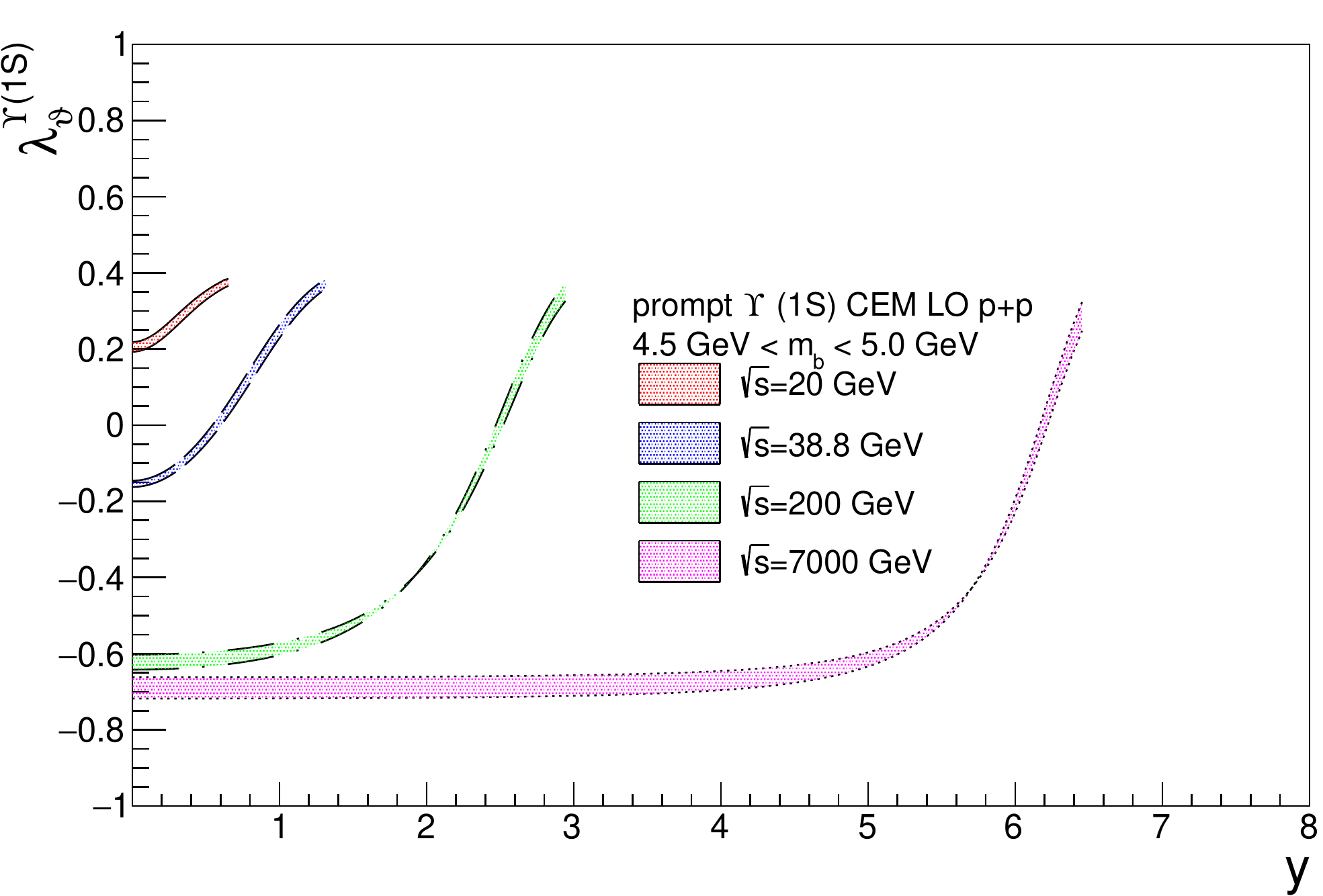}
\caption{\label{upsilon_lambda_y} The rapidity dependence of the polarization parameter $\lambda_\vartheta$ for production of prompt $\Upsilon$(1S) at $\sqrt{s}=20$~GeV (solid), $\sqrt{s}=38.8$~GeV (dashed), 200~GeV (dot-dashed), and 7000~GeV (dotted). The distributions are symmetric around $y=0$.}
\end{figure}

\subsection{Energy dependence of $\lambda_\vartheta$}

In this section, we compare the energy dependence of the polarization parameter $\lambda_\vartheta$ as a function of center of mass energy in $p+p$ collisions in Figs.~\ref{psi_lambda} and \ref{upsilon_lambda}.  The integration in Eq.~(\ref{cem_sigma}) for the direct production of each quarkonium state $Q$ is from the mass of the quarkonium state $M_Q$ to twice the mass of the lowest lying open heavy flavor hadron. The longitudinal to unpolarized ratios for the direct productions are then weighed to give the longitudinal to unpolarized ratio for the prompt production by Eqs.~(\ref{mix_psi}) and (\ref{mix_upsilon}) using parameters listed in Table~\ref{states}. The polarization parameters for prompt production and $J^P=1^-$ (S states) are then calculated using Eq.~(\ref{s_state_lambda}). The polarization parameter for direct production of $J^P=1^+$($\chi_1$P states) and $2^+$($\chi_2$P states) are calculated employing Eqs.~(\ref{chi_1_lambda}) and (\ref{chi_2_lambda}) respectively. The mass of the charm quark $m_c$ is varied around the base value 1.27~GeV from 1.2~GeV to 1.5~GeV while the mass of the bottom quark, $m_b$, is varied around the base value 4.75~GeV from 4.5~GeV to 5.0~GeV to construct the uncertainty bands shown in the figures.


\subsubsection{Direct production of $J/\psi$, $\psi$(2S), $\chi_{c2}$(1P), and prompt production of $J/\psi$}

In Fig.~\ref{psi_lambda}, the polarization paramters as a function of energy for direct production of the charmonium states below the hadron threshold and the prompt production of $J/\psi$ is presented. The integral over the pair invariant mass is assumed to be from $M_Q$ to $2m_{D^0}$ ($m_{D^0}=1.86$~GeV). We see that all direct production of $J/\psi$, $\chi_{c2}$(1P), and $\psi$(2S) is longitudinal for $\sqrt{s} > 20$~GeV. The prompt production of $J/\psi$ (bounded by blue filled solid curves in Fig.~\ref{psi_lambda}) is longitudinally polarized for $\sqrt{s} > 10$~GeV. Both direct and prompt productions becomes more longitudinal as $\sqrt{s}$ increases. The polarization of direct $\psi$(2S) is less longitudinal than that of direct $J/\psi$. This is because the improved CEM integrate from the mass of quarkonium to the hadron threshold, otherwise the direct $J/\psi$ and $\psi$(2S) results would be equal since the traditional CEM uses $2m_c$ for the lower limit of integration for all states.  The longitudinal to unpolarized fraction decreases as a function of $\sqrt{\hat{s}}$ for $J^P=1^-$ production. The hadron level longitudinal to unpolarized fraction will be smaller for direct $\psi$(2S) due to its larger mass.  Thus its polarization is less longitudinal.  Prompt $J/\psi$ production is dominated  by the S states and thus is longitudinally polarized. At $\sqrt{s} > 100$~GeV, the polarization parameter for prompt $J/\psi$ production saturates at $\lambda_\vartheta = -0.51^{+0.05}_{-0.16}$ while the polarization parameter for direct $J/\psi$ saturates at $\lambda_\vartheta = -0.61^{+0.07}_{-0.21}$.

The polarization parameter for direct $\chi_{c1}$ production is not shown in Fig.~\ref{psi_lambda} because the direct production yields only $J_z=\pm1$ by Eqs.~(\ref{chi_1_qq}) and (\ref{chi_1_gg}) and thus Eq.~(\ref{chi_1_lambda}) gives $\lambda_\vartheta = -1/3$.


\begin{figure}
\centering
\includegraphics[width=\columnwidth]{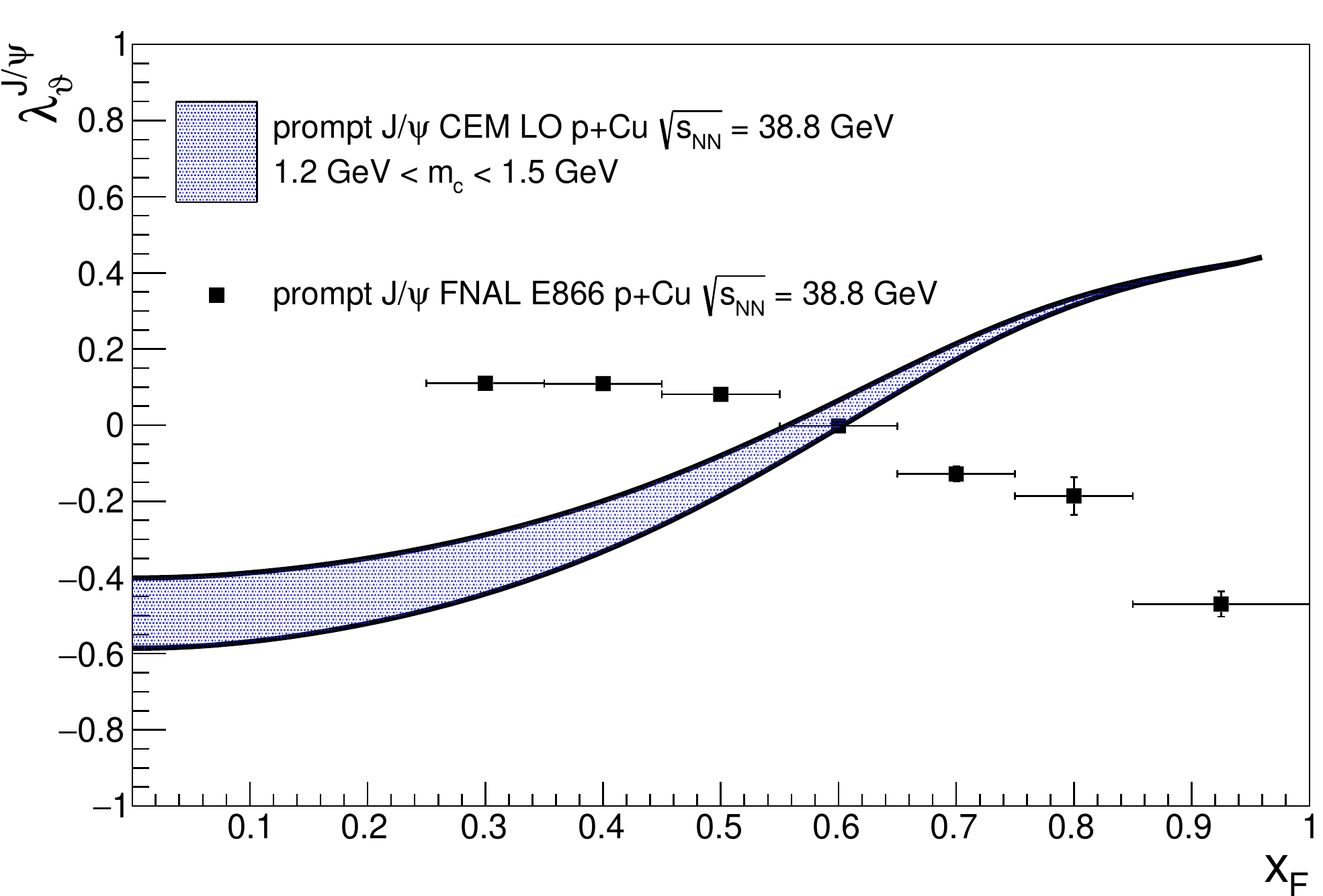}
\caption{\label{psi_lambda_xf} The $x_F$ dependence of the polarization parameter $\lambda_\vartheta$ for prompt production of $J/\psi$ in $p$+Cu collisions at $\sqrt{s_{NN}}=38.8$~GeV is compared to the E866/NuSea data \protect\cite{Chang:1999hd,Chang:2003rz}. The horizontal uncertainties are the experimental bin widths.}
\end{figure}

\subsubsection{Direct production of $\Upsilon$(1S), $\Upsilon$(2S), $\Upsilon$(3S), $\chi_{c2}$(1P), $\chi_{c2}$(2P), and prompt production of $\Upsilon$(1S)}

The results for direct production of the bottomonium states and prompt production of $\Upsilon$(1S) are shown in Fig.~\ref{upsilon_lambda}. Here, the integral over the pair invariant mass is assumed to be from $M_Q$ to $2m_{B^0}$ ($m_{B^0}=5.28$~GeV). For the more massive bottom quarks, direct production of $\Upsilon$(1S), $\Upsilon$(2S), and $\Upsilon$(3S) starts out transversely polarized for $\sqrt{s} < 34$~GeV. This is because $q\overline{q}\rightarrow Q\overline{Q}$ dominates the total cross section at these energies. As the $gg\rightarrow Q\overline{Q}$ contribution rises, the longitudinal fraction $R_\Upsilon$ increases and the direct production becomes longitudinal. As a result, the direct production of $\Upsilon$(1S), $\Upsilon$(2S), $\Upsilon$(3S), $\chi_{c2}$(1P), $\chi_{c2}$(2P), and prompt production production of $\Upsilon$(1S) becomes dominated by longitudinal polarization. Similar to charmonium production, the direct production of $\Upsilon$(1S) is mostly longitudinally polarized at collider energies, followed by $\Upsilon$(2S) and $\Upsilon$(3S) due to the increase in the lower limit of integration. However, for the case of bottomonium production, the longitudinal to unpolarized ratio at the parton level decreases slower as a function of $\sqrt{\hat{s}}$ in the integration range. This makes the bottomomium polarization relatively less sensitive to the mass of quark comparied to charmonium polarization. The polarization parameter for prompt $\Upsilon$(1S) saturates at $\lambda_\vartheta = -0.69^{+0.03}_{-0.02}$ while the polarization parameter for direct $\Upsilon$(1S) saturates at $\lambda_\vartheta = -0.91^{+0.04}_{-0.03}$ for $\sqrt{s}\sim 1$~TeV.  Note that the limit is lower for prompt $\Upsilon$(1S) than for prompt $J/\psi$ at the same energy.

Prompt production of $\Upsilon$(1S) is unpolarized ($\lambda_\vartheta = 0$) for $\sqrt{s} = 34$~GeV. The polarization parameters for direct $\chi_{b1}$(1P) and $\chi_{b1}$(2P) production are not shown in Fig.~\ref{upsilon_lambda} because direct production is only via $J_z=\pm1$ according to Eqs.~(\ref{chi_1_qq}) and (\ref{chi_1_gg}) and thus Eq.~(\ref{chi_1_lambda}) gives $\lambda_\vartheta = -1/3$.

\begin{figure}
\centering
\includegraphics[width=\columnwidth]{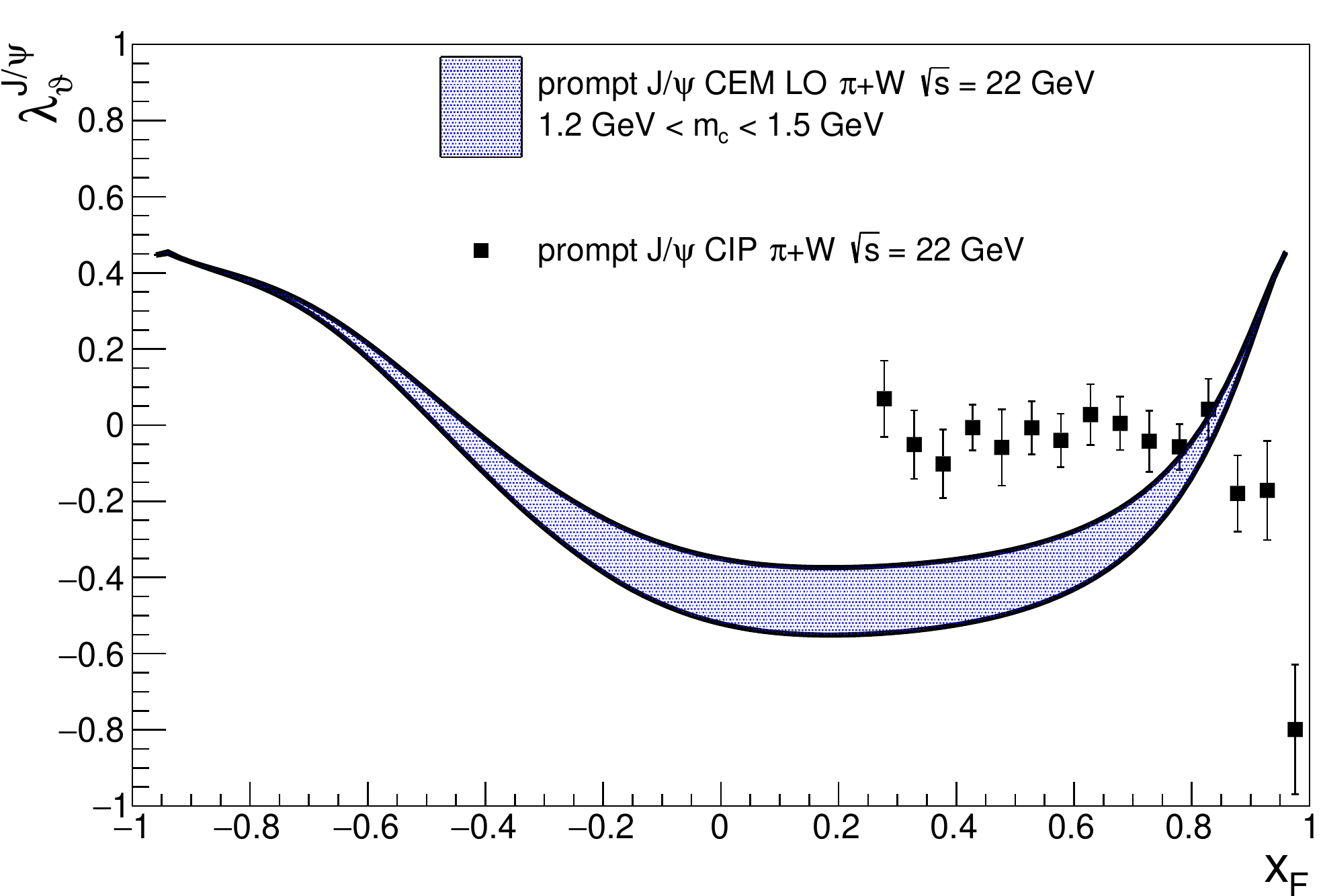}
\caption{\label{psi_lambda_xf_piW} The $x_F$ dependence of the polarization parameter $\lambda_\vartheta$ for prompt production of $J/\psi$ in $\pi$+W collisions at $\sqrt{s}=22$~GeV are compared to the CIP data \cite{Biino:1987qu}.}
\end{figure}


\subsection{Rapidity dependence of $\lambda_\vartheta$}

We now turn to the rapidity dependence of our result, shown in Figs.~\ref{psi_lambda_y} and \ref{upsilon_lambda_y}.  The direct production of each quarkonium state $Q$ is obtained by integrating Eq.~(\ref{cem_rapidity}) from the mass of the quarkonium state $M_Q$ to twice the mass of the lowest lying open heavy flavor hadron. The longitudinal to unpolarized ratios for the direct productions are then weighed to give the longitudinal to unpolarized ratio for the prompt production by Eqs.~(\ref{mix_psi}) and (\ref{mix_upsilon}) using the $c_Q$ values listed in Table~\ref{states}. The polarization parameters for prompt production is then found by Eq.~(\ref{s_state_lambda}). Four representative energies are chosen to illustrate. The lowest values, $\sqrt{s} = 20$ and 38.8~GeV were the highest available fixed-target energies at the CERN SPS for ion beams and the FNAL Tevatron for proton beams. The higher energies, $\sqrt{s} = 0.2$ and 7~TeV are energies available at the BNL RHIC and CERN LHC facilities. The results are presented for positive rapidity only because the rapidity distributions are symmetric around $y=0$ in $p+p$ collisions. Again, the charm quark mass $m_c$ is varied around 1.27~GeV from 1.2~GeV to 1.5~GeV while the bottom quark mass $m_b$ is varied around 4.75~GeV from 4.5~GeV to 5.0~GeV to construct the uncertainty bands.

\subsubsection{Direct production of $J/\psi$, $\psi$(2S), $\chi_{c2}$(1P), and prompt production of $J/\psi$}

The rapidity dependence of the polarization parameter for prompt $J/\psi$ is shown in Fig.~\ref{psi_lambda_y}. The results are given up to the kinematic limits of production. The polarization parameter is negative with a minimum at $y = 0$ and increases as $|y|$ increases, becoming positive at the kinematic limit. For the highest energies, where the longitudinal polarization has saturated in Fig.~\ref{psi_lambda}, the polarization parameter is flat over a wide range of rapidity.  The parameter remains negative as long as the $gg \rightarrow Q\overline{Q}$ contribution, with a significant longitudinal polarization, dominates production. As the phase space for charmonium production is approached, the $q \overline q \rightarrow Q \overline{Q}$ channel, predominantly transversely polarized, begins to dominate, causing the parameter to increase to a maximum of $\lambda_\vartheta \sim 0.4$.


\begin{figure}
\centering
\includegraphics[width=\columnwidth]{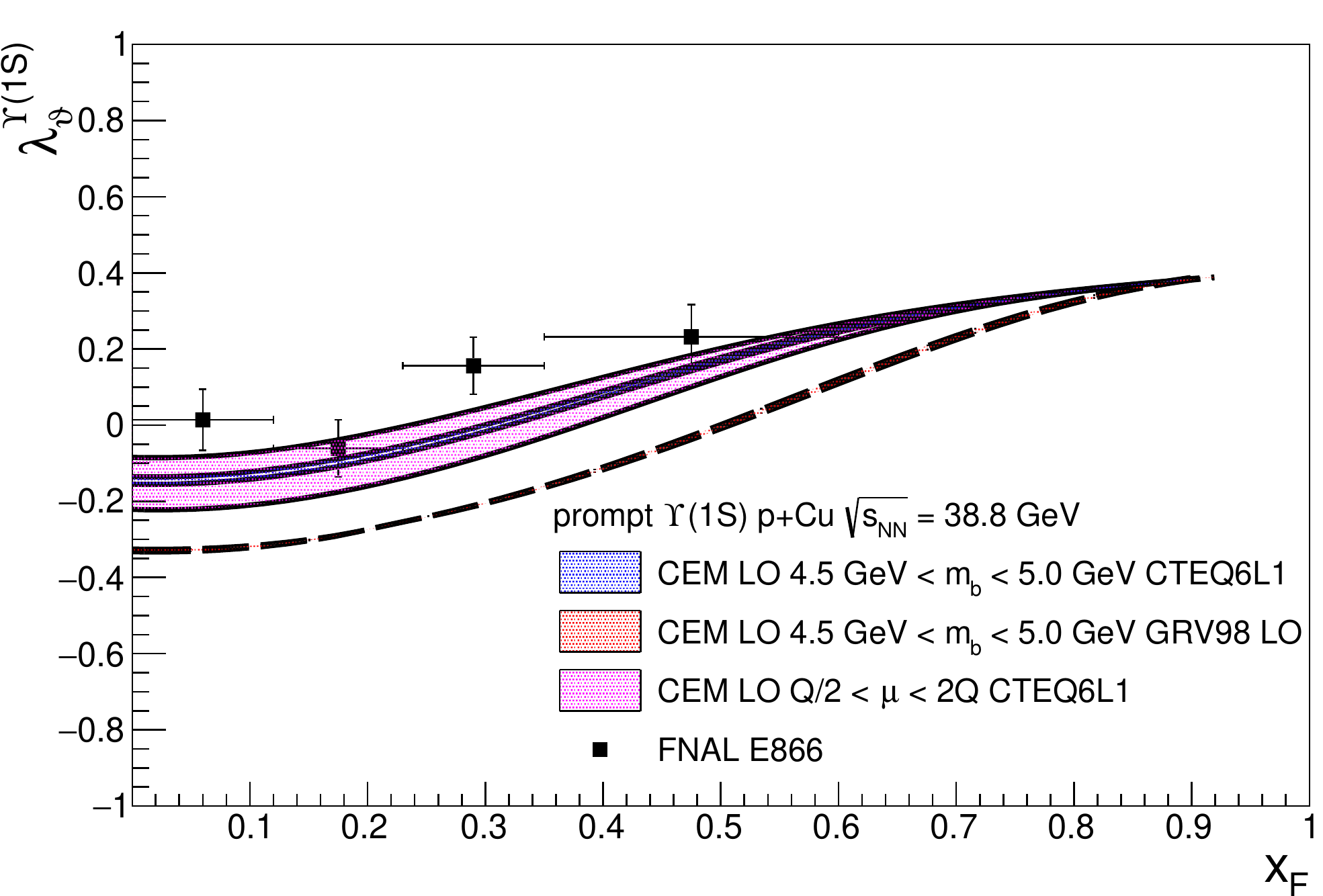}
\caption{\label{upsilon_lambda_xf} The $x_F$ dependence of the polarization parameter $\lambda_\vartheta$ for prompt production of $\Upsilon$(1S) in $p$+Cu collisions at $\sqrt{s}=38.8$~GeVusing CTEQ6L1 and varying $m_b$ (blue solid), GRV98 LO and varying $m_b$ (red dashed), CTEQ6L1 and varying $Q$ (magenta solid), and the data (box). The horizontal uncertainties on the E866/NuSea data \protect\cite{Brown:2000bz} are the bin widths.}
\end{figure}

\subsubsection{Direct production of $\Upsilon$(1S), $\Upsilon$(2S), $\Upsilon$(3S), $\chi_{c2}$(1P), $\chi_{c2}$(2P), and prompt production of $\Upsilon$(1S)}

The behavior of the prompt $\Upsilon$(1S) polarization parameter as a function of rapidity, shown in Fig.~\ref{upsilon_lambda_y}, is similar to that of prompt $J/\psi$.  The higher mass scale, however, reduces the kinematic range of the calculation. It also results in an unpolarized to slightly transverse polarization of prompt $\Upsilon$(1S) at fixed-target energies. At $\sqrt{s} = 20$~GeV, not far from production threshold, prompt $\Upsilon$(1S) is transversely polarized in the narrow rapidity range of production.


\subsection{Comparison to fixed-target data}

In this section, we compare our results as a function of longitudinal momentum fraction $x_F$ using Eq.~(\ref{cem_xf}) with the polarization parameters measured in fixed-target experiments. We compare our results to the results from the E866/NuSea Collaboration for the polarization of $J/\psi$ \cite{Chang:1999hd,Chang:2003rz} and $\Upsilon$(1S) \cite{Brown:2000bz} in $p$+Cu collisions at $\sqrt{s_{NN}}=38.8$~GeV as well as $J/\psi$ in $\pi$+W at $\sqrt{s}=22$~GeV by the CIP Collaboration \cite{Biino:1987qu}. We multiply the CTEQ6L1 PDFs by the central EPS09 \cite{Eskola:2010jh} nuclear modification to obtain the PDFs for Cu and W. We employ the GRS99 \cite{Gluck:1999xe} pion PDFs. The polarizations measured by the E866/Nusea Collaboration are made in Collins-Soper frame and the polarization measured by the CIP Collaboration is measured in Gottfried-Jackson frame. However, at leading order, the polarization axes in the helicity frame, the Collins-Soper frame, and the Gottfried-Jackson frame frame are coincident \cite{Faccioli:2010kd}.


\subsubsection{Prompt production of $J/\psi$ in $p$+Cu collisions at $\sqrt{s_{NN}}=38.8$~GeV}

We compare our polarization predictions for prompt production of $J/\psi$ in $p$+Cu collisions at $\sqrt{s}=38.8$~GeV as a function of $x_F$ on the results measured by E866/NuSea Collaboration \cite{Chang:1999hd,Chang:2003rz} and is shown in Fig.~\ref{psi_lambda_xf}. Since the $x_F$ dependence is nearly symmetric around $x_F=0$, the result is presented for positive $x_F$ only. Both $J/\psi$ and $\psi$(2S) are included in the experimental results but only about 1\% of the contribution comes from the $\psi$(2S). Our result is longitudinal at small values of $x_F$ and becomes transverse at large $x_F$. The experimental results disagree with ours since the polarization parameter measured decreases as a function of $x_F$. Our $x_F$ integrated prediction is $\lambda_\vartheta = -0.41^{+0.05}_{-0.13}$ while the experimental result reports $\lambda_\vartheta = 0.069 \pm 0.004$.

\subsubsection{Prompt production of $J/\psi$ in $\pi$+W collisions at $\sqrt{s}=20$~GeV}

We compare our polarization predictions for prompt production of $J/\psi$ in $\pi$+W collisions at $\sqrt{s}=20$~GeV as a function of $x_F$ to the measurement by the CIP Collaboration \cite{Biino:1987qu} in Fig.~\ref{psi_lambda_xf_piW}. The $x_F$ dependence is not symmetric around $x_F = 0$ in this case due to the difference in the high $x$ behavior of the pion PDFs relative to that of the proton PDFs. Therefore the result is shown over all $x_F$. We note that the polarization predictions differ slightly in $\pi$+W collisions at $\sqrt{s}=20$~GeV than in $p$+Cu collisions at $\sqrt{s_{NN}}=38.8$~GeV. The polarization at $x_F=0$ is less longitudinal in $\pi$+W collisions although the trend is similar: longitudinal polarization at small values of $x_F$ and transverse at large $x_F$. The experimental results disagree with ours since the polarization parameter measured is near unpolarized as a function of $x_F$ except for the last $x_F$ bin. However, our prediction reaches a better agreement with data in $\pi$+W compared to $p$+Cu in terms of the behavior as a function of $x_F$. Our result predicts in the region of low to mid positive $x_F$, $J/\psi$ is produced with a relatively constant moderate longitudinal polarization. Our $x_F$ integrated prediction is $\lambda_\vartheta = -0.42^{+0.05}_{-0.13}$ while the experiment reports $\lambda_\vartheta = -0.02 \pm 0.06$.


\subsubsection{Prompt production of $\Upsilon$(1S) in $p$+Cu collisions at $\sqrt{s_{NN}}=38.8$~GeV}

We now turn to the $x_F$ dependence of the polarization parameter in prompt $\Upsilon$(1S) production. We compare our polarization predictions for prompt production of $\Upsilon$(1S) in $p$+Cu collisions at $\sqrt{s}=38.8$~GeV to the results measured by E866/NuSea Collaboration \cite{Brown:2000bz} in Fig.~\ref{upsilon_lambda_xf}. This is the lowest energy at which $\Upsilon$(1S) polarization has been measured. Our results is slightly longitudinal at small values of $x_F$ and becomes slightly transverse at large $x_F$. Our results are comparable to the data since both the predicted and measured polarization parameters increase as function of $x_F$. Our result is consistent with the $\sim 0$ polarization measured by the E866/NuSea Collaboration. The measured polarization for $\Upsilon$(1S) independent of $x_F$ is $\lambda_\vartheta = 0.07 \pm 0.04$ while our prediction is $\lambda_\vartheta = -0.06 \pm 0.01$.


\subsection{Polarization predictions for prompt production of $J/\psi$ and $\Upsilon$(1S) in $p$+Pb collisions at fixed-target energies at the LHC}

\begin{figure}
\centering
\includegraphics[width=\columnwidth]{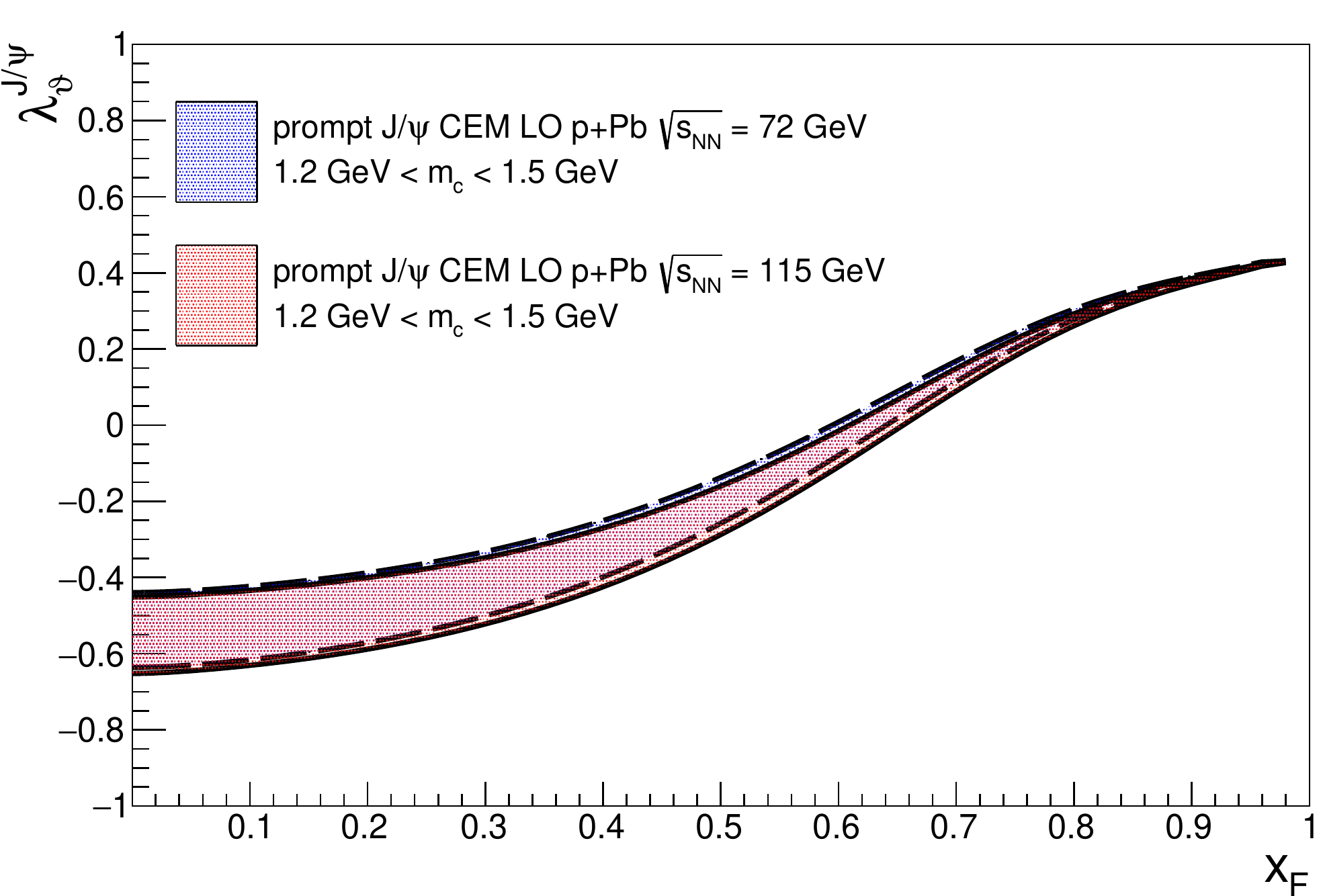}
\caption{\label{psi_lambda_xf_pPb} The $x_F$ dependence of the polarization parameter $\lambda_\vartheta$ for production of $J/\psi$ in $p$+Pb at $\sqrt{s_{NN}}=72$~GeV (blue dashed) and 115~GeV (red solid).}
\end{figure}

In this section, we present our polarization predictions for prompt production of $J/\psi$ and $\Upsilon$(1S) as a function of $x_F$ using Eq.~(\ref{cem_xf}) for $p$+Pb fixed-target interactions at the LHC. The polarization predictions are presented for $\sqrt{s_{NN}}=72$~GeV and 115~GeV, the center of mass energies for a lead beam on a proton target and a proton beam on a lead target respectively. Since the $x_F$ dependence is nearly symmetric around $x_F=0$, the results are only presented for postitive $x_F$. We again multiply the CTEQ6L1 PDFs by the central EPS09 nuclear modification to obtain the lead PDFs. Also, since our predictions are calculated at leading order, they are frame independent.


\begin{figure}
\centering
\includegraphics[width=\columnwidth]{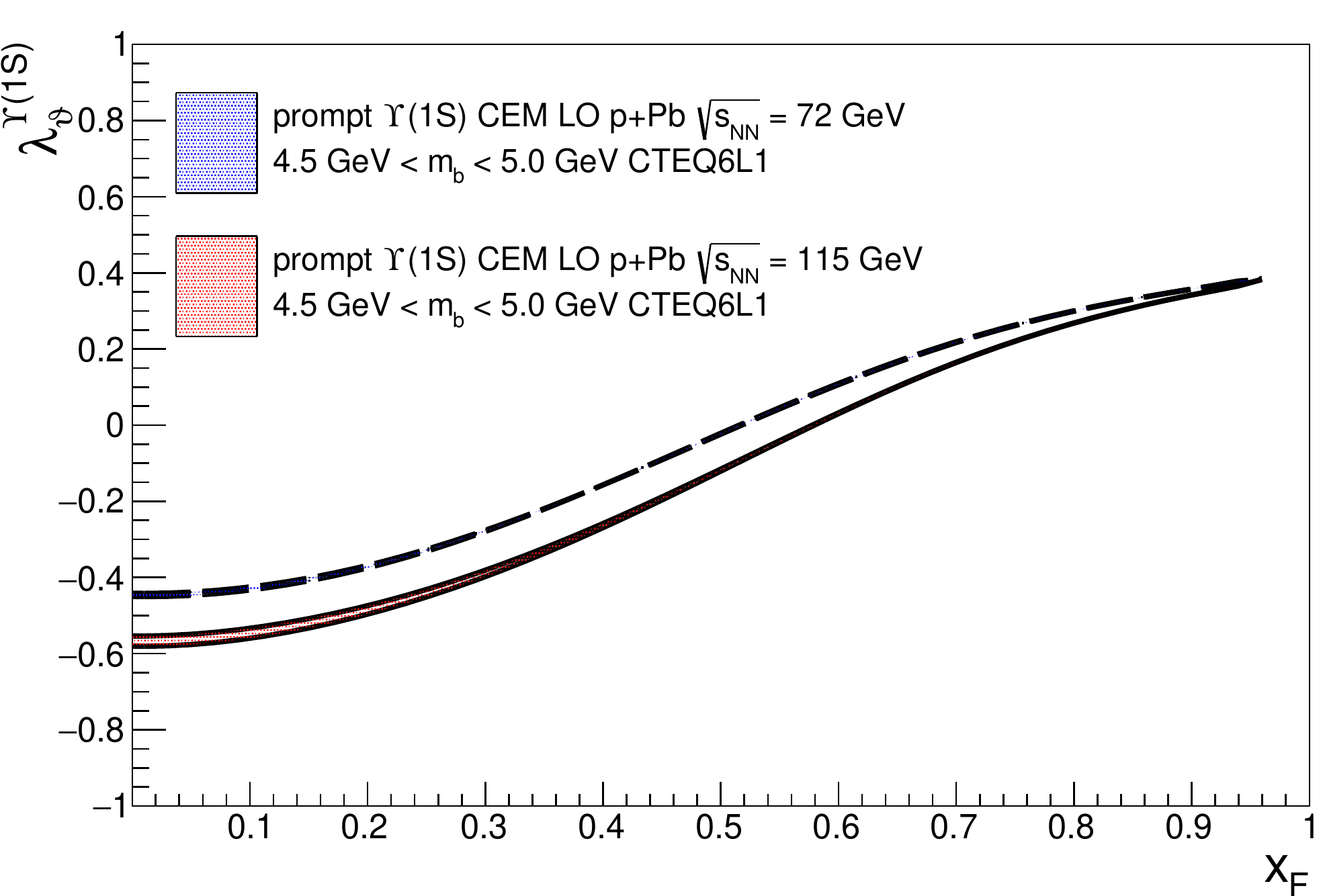}
\caption{\label{upsilon_lambda_xf_pPb} The $x_F$ dependence of the polarization parameter $\lambda_\vartheta$ for production of $\Upsilon$(1S) in $p$+Pb at $\sqrt{s_{NN}}=72$~GeV (blue dashed), 115~GeV (red solid).}
\end{figure}

\subsubsection{Prompt $J/\psi$ production at the LHC}

We present our polarization prediction for prompt $J/\psi$ production in $p$+Pb interactions at $\sqrt{s_{NN}}=72$~GeV and 115~GeV as a function of $x_F$ in Fig.~\ref{psi_lambda_xf_pPb}. The longitudinal polarization already starts to saturate at these energies for prompt $J/\psi$ production as presented in Fig.~\ref{psi_lambda}. Therefore, the polarization for prompt $J/\psi$ production at these energies are very similar. The polarization is longitudinal at small $x_F$ and becomes transverse at large $x_F$. Our $x_F$-integrated prediction is $\lambda_\vartheta = -0.46^{+0.04}_{-0.15}$ at $\sqrt{s_{NN}}=72$~GeV and $\lambda_\vartheta = -0.46^{+0.03}_{-0.17}$ at $\sqrt{s_{NN}}=115$~GeV. 


\subsubsection{Prompt $\Upsilon$(1S) production at the LHC}

The prediction for polarization of prompt $\Upsilon$(1S) production in $p$+Pb collisions at $\sqrt{s_{NN}}=72$~GeV and 115~GeV is given as a function of $x_F$ in Fig.~\ref{upsilon_lambda_xf_pPb}. Because of the higher mass scale, the longitudinal polarization is not saturated at these energies for prompt $\Upsilon$(1S) production. Therefore, the polarization for prompt $\Upsilon$(1S) prodcution at these energies are different. The behavior of the polarization at both energies are similar. Prompt $\Upsilon$(1S) is longitudinal at small $x_F$ and becomes transvere at large $x_F$. However, the polarization at $\sqrt{s_{NN}}=115$~GeV is more lonitudinal. Our $x_F$ integrated prediction is $-0.367^{+0.002}_{-0.001}$ at $\sqrt{s_{NN}}=72$~GeV and $-0.51^{+0.01}_{-0.01}$ at $\sqrt{s_{NN}}=115$~GeV.   


\subsection{Sensitivity to the proton PDFs}
\label{PDF}

We have tested the sensitivity of our results to the choice of PDFs used in the calculation. Since few new LO proton PDFs are currently available, we compare our CTEQ6L1 results with calculations using the older GRV98 LO \citep{Gluck:1998xa} set.  We can expect the ratio to be the most sensitive to the choice of proton PDF because the PDFs can change the balance of $gg$ to $q\overline{q}$ production, especially at lower $\sqrt{s}$ where the $x$ values probed by the calculations are large, $x \sim 0.1$.  In particular, the prompt production of $\Upsilon$(1S) at $\sqrt{s} = 20$~GeV is most likely to be sensitive to the choice of PDF since the $q\overline{q}$ contribution is large at this energy.  The results should, on the other hand, be relatively insensitive to the chosen mass and scale values since these do not strongly affect the relative contributions of $gg$ and $q\overline{q}$.

\begin{table}
\caption{\label{feed_down_sensitivity}Values of $c_Q$ used to test the sensitivity of our results to the feed down ratios. Base on the uncertainty in $c_Q$, $c_Q^\prime$ is used assuming the promptly produced 1S states comprise less directly produced 1S states, and $c_Q^{\prime \prime}$ is used assuming the promptly produced 1S states comprise more directly produced 1S states,}
\begin{ruledtabular}
\begin{tabular}{cccc}
$Q$ & $c_{Q}^\prime$ & $c_{Q}$ & $c_{Q}^{\prime \prime}$\\
\hline
$J/\psi$ & 0.59 & 0.62$\pm$0.04 & 0.65  \\
$\psi$(2S) & 0.09 & 0.08$\pm$0.02 & 0.07 \\
$\chi_{c1}$(1P) & 0.17 & 0.16$\pm$0.04 & 0.15 \\
$\chi_{c2}$(1P) & 0.15 & 0.14$\pm$0.04 & 0.13 \\
$\Upsilon$(1S) & 0.43 & 0.52$\pm$0.09 & 0.61 \\
$\Upsilon$(2S) & 0.12 & 0.1$\pm$0.03 & 0.08 \\
$\Upsilon$(3S) & 0.03 & 0.02$\pm$0.005 & 0.01 \\
$\chi_{b1}$(1P) & 0.145 & 0.13$\pm$0.035 & 0.115 \\
$\chi_{b2}$(1P) & 0.145 & 0.13$\pm$0.035 & 0.115 \\
$\chi_{b1}$(2P) & 0.065 & 0.05$\pm$0.025 & 0.035 \\
$\chi_{b2}$(2P) & 0.065 & 0.05$\pm$0.025 & 0.035 \\
\end{tabular}
\end{ruledtabular}
\end{table}

This is indeed the case, for prompt $\Upsilon$(1S) production at $\sqrt{s} = 20$~GeV, close to the production threshold, the largest difference in the longitudinal ratio for the two PDF sets is 15\% at $y=0$, making a difference in the polarization parameter, $\lambda_\vartheta$ of 0.35 around the unpolarized region. The sensitivity arises because the $gg$ contribution in the prompt productions of the S states are predominantly produced with $J_z = 0$ while the $q\overline{q}$ contribution is primarily produced with $J_z = \pm 1$.  By $\sqrt{s} = 38.8$~GeV, the difference in the results is reduced to 9\%, making a difference in $\lambda_\vartheta$ of 0.18 around the slightly longitudinal region. The $x_F$ dependence of prompt $\Upsilon$(1S) polarization using GRV98 LO is also shown along with the prediction using CTEQ6L1 in Fig.~\ref{upsilon_lambda_xf}. The prediction using GRV98 LO is more longitudinal compared to the prediction using CTEQ6L1. At collider energies, the difference is negligible. Since the $gg$ contribution is dominant for $J/\psi$ already at $\sqrt{s} = 20$~GeV, the prompt $J/\psi$ production polarization is essentially independent of the choice of proton PDF.  Thus, away from production threshold, the results are robust with respect to the choice of PDF.


\subsection{Sensitivity to factorization scale}

We have tested the sensitivity of our results to the factorization scale, $\mu$. We varied the factorization scale for prompt $J/\psi$ and $\Upsilon$(1S) in the range: $Q/2 \leq \mu \leq 2Q$ while keeping the renormalization scale the same. We have found the longitudinal to unpolarized fractions $R^{J_z=0}_{J/\psi}$ and $R^{J_z=0}_{\Upsilon \rm{(1S)}}$ are hardly changed in the range of $\mu$ varied at high energies where the polarization is saturated. The ratio for each directly produced charmonium $R^{J_z=0}_{\psi}$ is changed by $\sim 0.01$ while $R^{J_z=0}_{\Upsilon}$ is changed by $\sim 0.001$ for each directly produced botommonium. We note that each indiviual polarized production cross section is affected by the variation in factorization scale. But at high energies, the production is dominated by the gluon fusion processes. Therefore, the polarization, which depends on the longitudinal to unpolarized ratio, is not sensitive to the factorization scale. 

However, at fixed-target energies, where gluon fusion does not yet dominate production, the polarization is affected by the variation in the factorization scale. Indeed, the uncertainty bands for prompt $\Upsilon$(1S) polarization due to varying the factorization scale is wider than that for varying the bottom quark mass at fixed-target energies. We also present the polarization of prompt $\Upsilon$(1S) by varying the factorization scale at $\sqrt{s_{NN}}=38.8$~GeV in Fig.~\ref{upsilon_lambda_xf}. At $\sqrt{s_{NN}}=38.8$~GeV, the uncertainty on the polarization of prompt $\Upsilon$(1S) due to changing the factorization scale is $-0.05^{+0.05}_{-0.08}$, slightly closer to the measured polarization by the E866 Collaboration than that from varying the bottom quark mass.  However, the uncertainty band due to factorization scale variation for prompt $J/\psi$ is smaller than that due to changing $m_c$ for all energies.  This is because the polarization of prompt $J/\psi$ saturate at a lower energy compared to prompt $\Upsilon$(1S).


\subsection{Sensitivity to feed down ratios}

We have tested the sensitivity of our results to the feed down ratios we use in our calculations \cite{Digal:2001ue}. Since the prompt production of $J/\psi$ and $\Upsilon$(1S) are dominated by direct $J/\psi$ and direct $\Upsilon$(1S) respectively, we vary the feed down ratio by changing the relative contribution by direct $J/\psi$ and direct $\Upsilon$(1S) to other states. That is when $c_{J/\psi}$ increase, all other $c_{\psi}$ decreases and vice versa, similarly for $c_{\Upsilon \rm (1S)}$ and other $c_{\Upsilon}$. Using the base values of $c_\psi$ and $c_\Upsilon$ in Table~\ref{states} and the reported uncertainty, we vary the feed down ratios as given in Table~\ref{feed_down_sensitivity}.  Considering only the variation of the feed down ratios, the uncertainty on the polarization parameter for prompt $J/\psi$ production at $\sqrt{s}=7$~TeV is $\lambda_\vartheta = -0.51\pm0.01$. These uncertainties are much smaller than those due to charm quark mass variation.  The uncertainty on the polarization parameter for prompt $\Upsilon$(1S) at $\sqrt{s}=7$~TeV is $\lambda_\vartheta = -0.69^{+0.03}_{-0.04}$ due to changing $c_\Upsilon$. These uncertainties are very similar to those due to varying $m_b$.


\section{Conclusions}

We have presented the energy and rapidity dependence of the polarization of prompt $J/\psi$ and $\Upsilon$(1S) production in $p+p$ collisions in the Color Evaporation Model. We compare the $x_F$ dependence to experimental results in $p$+Cu and $\pi$+W collisions at fixed-target energies. We also present our polarization predictions as a function of $x_F$ for fixed-target experiments at the LHC. We find prompt $J/\psi$ and $\Upsilon$(1S) production to be longitudinally polarized, saturating at energies far above the $Q \overline Q$ production threshold, with $\lambda_\vartheta^{J/\psi}=-0.51^{+0.05}_{-0.16}$ and $\lambda_\vartheta^{\Upsilon \text{(1S)}}=-0.69^{+0.03}_{-0.02}$. We find the prompt $J/\psi$ and $\Upsilon$(1S) polarization to be longitudinal around central rapidity while the polarization becomes transverse as the kinematic limits of the calculation, where $q \overline q$ production is dominant, are approached.

Since our calculation is leading order, we cannot yet calculate the $p_T$ dependence of quarkonium polarization.  This will be addressed in a future publication.


\section{Acknowledgemets}
We thank F. Yuan for valuable discussions throughout this work. This work was performed under the auspices of the U.S. Department of Energy by Lawrence Livermore National Laboratory under Contract DE-AC52-07NA27344 and supported by the U.S. Department of Energy, Office of Science, Office of Nuclear Physics (Nuclear Theory) under contract number DE-SC-0004014.


\end{document}